%% file: ThermalPaper.tex
\documentclass[final,5p,times]{elsarticle}

\usepackage[english]{babel}     
\usepackage[T1]{fontenc}        
\usepackage[utf8]{inputenc}     

\usepackage{amsmath,amsfonts,amssymb,bm} 
\usepackage[per-mode = symbol]{siunitx}
\usepackage[activate={true,nocompatibility},final,tracking=true,kerning=true,spacing=true,factor=1100,stretch=10,shrink=10]{microtype}
\microtypecontext{spacing=nonfrench}

\usepackage{graphicx}   
\usepackage{caption}
\usepackage{subcaption} 

\usepackage[section]{placeins}  

\usepackage{epstopdf}   

\usepackage{booktabs}   
\usepackage{multirow}   
\usepackage{threeparttable} 

\usepackage[switch]{lineno}
\modulolinenumbers[5]

\usepackage[hyphens]{url}
\usepackage[pdftex]{hyperref}       
\hypersetup{colorlinks          = true,         
            allcolors           = black,        
            bookmarksnumbered   = true,         
            bookmarksopen       = true,         
            breaklinks          = true,         
            pdfdisplaydoctitle  = true,         
            pdfpagelayout       = OneColumn,    
            pdfstartview        = FitH,         
            pdfpagemode         = UseNone,      
            pdftitle            = {Thermal analysis of the electronics of a CubeSat mission},
            pdfauthor           = {Andre Guerra, Diego Nodar, Ricardo Tubo},
            pdfsubject          = {Article on a thermal analysis of a group of PCBs and their components for a CubeSat mission.}
} 

\journal{Applied Thermal Engineering}

\usepackage{xcolor}         

\usepackage[color=blue!20!white,linecolor=red,figcolor=white,textwidth=3cm,colorinlistoftodos,%
]{todonotes}                                                

\begin{document}


\begin{frontmatter} 
    \title{Thermal analysis of the electronics of a CubeSat mission}
    
    \author[FCUP]{Andr\'{e} G. C. Guerra\corref{cor1}}
    \ead{aguerra@fc.up.pt}
    
    \author[VIGO]{Diego Nodar-L\'{o}pez}
    \ead{diego.nodar@space.uvigo.es}
    
    \author[BOLOGNA]{Ricardo Tub\'{o}-Pardavila}
    \ead{ricardo.tubio@unibo.it}
    
    \cortext[cor1]{Corresponding author}
    
    \address[FCUP]{Departamento de F\'{i}sica e Astronomia, Centro de F\'{i}sica do Porto, Faculdade de Ci\^{e}ncias, Universidade do Porto, Portugal} %
    
    \address[VIGO]{Escola de Enxe\~{n}ar\'{i}a de Telecomunicaci\'{o}n, Universidade de Vigo, Espa\~{n}a}

	\address[BOLOGNA]{Engineering and Architecture - Campus of Forli, University of Bologna, Italy}

    \begin{abstract}
    	CubeSat satellites are small cube-shaped space platforms that are very popular among academic, research and start-up communities mostly due to their reduced mission costs and short development time.
        This comes with an easy access to space, which enables high risk missions that would not be feasible otherwise.
        Thus, most of the subsystems available for these satellites have been adapted to a rather fast development methodology.
        However, thermal control engineering is still somewhat overlooked due to its complexity and long-term development characteristics.
        Moreover, since CubeSat projects often rely in commercial off-the-shelf components, it is necessary to understand their necessities in terms of thermal control.
        The main reason for this is because these components are not originally designed to operate in the space environment, and may degrade faster when operated either under too hot or too cold conditions. 
        In this context, this article presents the results of several thermal analyses and tests conducted over power amplifiers and field-programmable gate array devices, commonly used in CubeSat subsystems and payloads.
        Passive thermal control methods, mostly based on thermal straps, are explored as potential thermal control solutions due to their low cost and development constraints.
    \end{abstract}

    
\end{frontmatter}


\input{1_Intro}

\input{2_GeometricModel}

\input{3_Analysis}

\input{4_Results}

\input{6_Conclusion}


\section*{Acknowledgements}

The work of Andr\'{e} Guerra is supported by the Funda{\c c}\~ao para a Ci\^encia e a Tecnologia (Portuguese Agency for Research) fellowship PD/BD/113536/2015.
The authors would like to thank Isabel Perez Grande for her suggestions about this paper.
Andr\'{e} Guerra would also like to thank his PhD supervisor, Orfeu Bertolami, for discussions and suggestions.

\bibliography{ThermalPaper_BIB}

%

\end{document}

%% file: 1_Intro.tex
%
%
%

\section{Introduction}
\label{sec:Introduction}

The thermal control subsystem is one of the critical systems of almost any spacecraft, as it is responsible to guarantee that all the spacecraft's temperature is within a predefined range.
This range is dictated by requirements and constrains of the individual spacecraft components, considering all operational modes and environments that the spacecraft might be exposed to.
Besides respecting a given range of temperatures to operate in, the spacecraft should also limit the temperature gradients.
Otherwise, it could lead to reduce efficiencies and/or lifetime of the components, equipment malfunction, structure deformations or even total mission failure~\cite{Larson1999}.
Therefore, the thermal engineers must determine the operational conditions and propose solutions to control the temperature of the spacecraft.

In CubeSat technology (cube-shaped small satellites with mass ranging from \SIrange{1}{12}{kg}), thermal design is usually overlooked.
These spacecraft are characterised by having a minimum volume of \SI{1}{litre}, codenamed as 1U (1-unit CubeSat), although bigger configurations (using ``U'' as the base volume) are also allowed, including a 2U (2-units), a 3U (3-units) and a 3U-XL (which is a 3U with an additional tuna-can size volume)~\cite{TheCubeSatProgram2015}.
Moreover, CubeSats heavily rely on commercial off-the-shelf (COTS) electronic devices for most spacecraft subsystems, taking high performance components into space in very short development cycles.
The use of COTS is also leveraged by their light weight and capabilities, apart from their lower cost.
However, most of these COTS were designed to operate in a completely different environment, and placing them in space can have unforeseen implications.
This is especially relevant considering that the small satellite market, where CubeSats are included, has grown 205\% from 2016 to 2017, and more than 263 small satellites are expected to be launched in 2018~\cite{SpaceWorksEnterprisesInc2018}.

With this work we propose to study the temperature distribution of a planned CubeSat mission, including de-rating requirements like, for example, maximum joint temperature.
This CubeSat mission is set to be launched on 2018, and it will carry a field-programmable gate array (FPGA) device, among several other electronic devices, distributed over two PCBs (printed circuit boards), as described in section~\ref{sec:GeometricModel}.
Afterwards, the critical cases and major assumptions that have to be made are assessed, as explained in section~\ref{sec:Analysis}.
Safety margins had to be applied following standard recommendations (Ref.~\cite{ECSS_ThermalControl_2008}), since thermal design is bound to have numerous uncertainties, including contact conductances definitions, insulation characteristics, and methodologies limitations~\cite{Gilmore2003,Meseguer2012}.
The most relevant results are discussed in section~\ref{sec:Results}, while the conclusions and future actions are given in section~\ref{sec:Conclusion} and section~\ref{sec:Future_Work}.

The computational methodology used for the work described here is based on a finite element method (FEM) analysis and the Icepak software, from ANSYS (Ref.~\cite{ANSYS2016}).
The FEM methodology has several advantages.
For example, the mesh generated (which Icepak can do automatically), gives a temperature distribution over the components of the geometric model created.
The software has its own advantages, including having standard electronic packages, the easiness to add connections between components, or even the automatic view factor computation, for instance.
Nevertheless, there are alternative methods, like the lumped-parameter representation~\cite{Meseguer2012}, or others that have been used on the Pioneer, Cassini and New Horizon probes, for estimating the residual accelerations generated by the spacecraft thermal emissions~\cite{Francisco2012,Bertolami2014,Guerra2017}.


%% file: 2_GeometricModel.tex
%
%
%

\section{Geometric Model}
\label{sec:GeometricModel}

During the preliminary design phase, and in particular when using concurrent engineering approaches~\cite{ECSS_EngDesign_2010}, the design is highly iterative with constant changes.
It is thus important to build the required geometric models in a parametric fashion.
Although we only present here the final version, there is a small discussion of how this model was reached in section~\ref{sec:Analysis}.
Since the objective was to study the temperature of just a few components, some simplifications were made to the design.

\begin{figure}[!htb]
    \centering
    \includegraphics[width=0.9\columnwidth]{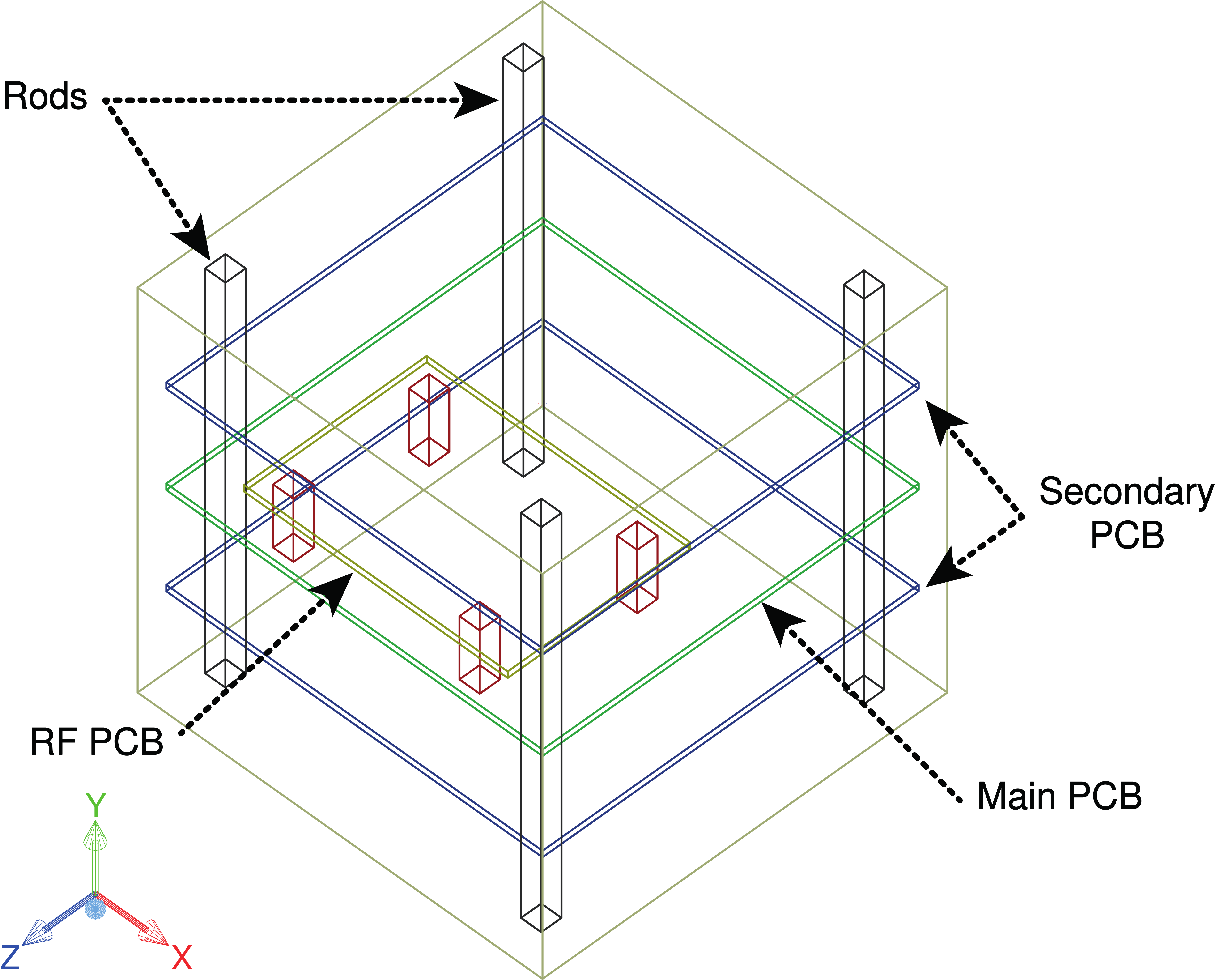}
    \caption{Basic model of the CubeSat structure, with the major components labelled.}
    \label{fig:BasicStructure}
\end{figure}

The geometry of the CubeSat is modelled as a hollow \SI[allow-number-unit-breaks=true]{10 x 10 x 10}{cm} cube, as per Ref.~\cite{TheCubeSatProgram2015}, with the electronic PCB boards hosted in an internal stack and the faces of the cube being modelled as walls.
The following assumptions have been made in order to simplify the thermal model:
\begin{enumerate}
    \item The outer face of the walls of the cube radiate to outer space and are modelled as having an average emissivity $\epsilon = 0.78$ and an absorptivity $\alpha = 0.37$.
    \item Internal components are thermally coupled with radiative interfaces only with the inner walls of the cube.
\end{enumerate}
A simplified schematic of this model is depicted in Figure~\ref{fig:BasicStructure}.

The internal stack has been modelled to have three major PCBs, plus a smaller one (the RF PCB) linked to the middle one.
Each of these PCBs is made of ten layers of Cu-Pure (which has a density of $\rho = \SI{8933}{kg/m^3}$, a conductivity $k$ equal to \SI{387.6}{W/m.K}, and a specific heat $C$ equal to \SI{195}{J/kg.K}, for a temperature of $T=\SI{77}{\celsius}$~\cite{ANSYS2016}).
These layers are stacked inside a FR-4 substrate ($\rho = \SI{1250}{kg/m^3}$, $C = \SI{1300}{J/kg.K}$ and $k = \SI{0.35}{W/m.K}$~\cite{ANSYS2016}) giving a total thickness of \SI{1.6}{mm}.
For thermal radiation exchange calculations we have to use the emissivity and the absorptance of FR-4, which is $\epsilon = 0.9$ and $\alpha = 0.9$ respectively.
Although all PCBs have the same interior structure, only the main one and the smaller RF PCB have individual electronic components.
Table~\ref{tab:PCBs_PosDim} summarises the dimensions and position of the PCBs.

\begin{table}[htbp]
    \footnotesize
    \centering
    \caption{PCBs position and dimensions inside the 1U CubeSat, with all values expressed in \si{cm}.}
    \label{tab:PCBs_PosDim}
    \begin{tabular}{cccccc}
        \toprule
            \multirow{2}[0]{*}{Component}
                & \multicolumn{2}{c}{Dimensions}    & \multicolumn{3}{c}{Position} \\
        \cmidrule{2-6}
                & Width & Length                    & X     & Y     & Z \\
        \midrule
            Main PCB 
                & \multirow{3}[0]{*}{9.29}  & \multirow{3}[0]{*}{9.29}
                                                    & 0.36  & 5.00  & 0.36 \\
            Secondary PCB Bottom 
                &       &                           & 0.36  & 2.50  & 0.36 \\
            Secondary PCB Top 
                &       &                           & 0.36  & 7.50  & 0.36 \\[1ex]
            RF PCB 
                & 6.51  & 4.51                      & 2.05  & 6.00  & 4.91 \\
        \bottomrule
    \end{tabular}
\end{table}

To support the PCBs there are four main rods in each corner, attached to the spacecraft walls at their endings.
The rods are made of aluminium, with a density of $\rho = \SI{2710}{kg/m^3}$, a conductivity $k = \SI{218}{W/m.K}$, a specific heat $C = \SI{900}{J/kg.K}$, an absorptance of $\alpha = 0.4$ and an emissivity $\epsilon = 0.09$~\cite{ANSYS2016}.
Besides these main ones, there are also four small rods that hold the RF PCB (using the main one as a base).
All of these rods dimensions and positions are shown in Table~\ref{tab:Rods_PosDim}.

\begin{table}[htbp]
    \scriptsize
    \setlength\tabcolsep{4pt}
    \centering
    \caption{Dimensions and positions of all the rods, with values expressed in \si{cm}.}
    \label{tab:Rods_PosDim}
    \begin{tabular}{ccccccc}
        \toprule
            \multirow{2}[0]{*}{Component}
                    & \multicolumn{3}{c}{Dimensions}
                        & \multicolumn{3}{c}{Position} \\
        \cmidrule{2-7}
            
                    & Outer Radius & Inner Radius & Height
                        & X     & Y     & Z \\
        \midrule
            Rod 1   & \multirow{4}[0]{*}{0.25} & \multirow{4}[0]{*}{0.15} & \multirow{4}[0]{*}{10.00} 
                        & 0.50  & \multirow{4}[0]{*}{0.00} & 0.97 \\
            Rod 2   &       &       &       
                        & 8.67  &       & 0.73 \\
            Rod 3   &       &       &       
                        & 8.67  &       & 8.70 \\
            Rod 4   &       &       &       
                        & 0.50  &       & 8.33 \\[1ex]
            Secondary Rod 1 
                    & \multirow{4}[0]{*}{0.25} & \multirow{4}[0]{*}{0.15} & \multirow{4}[0]{*}{1.56} 
                        & 2.54  & \multirow{4}[0]{*}{4.80} & 5.34 \\
            Secondary Rod 2 
                    &       &       &       
                        & 7.68  &       & 5.34 \\
            Secondary Rod 3 
                    &       &       &       
                        & 7.14  &       & 8.69 \\
            Secondary Rod 4 
                    &       &       &       
                        & 2.54  &       & 8.69 \\
        \bottomrule
    \end{tabular}
\end{table}

Each rod is composed of an inner tube, which is basically a threaded rod, and hollow spacers (meant to secure the PCBs in their positions), even though they are represented as a single piece in Figure~\ref{fig:BasicStructure}.
This was done to simplify the modelling of each rod, instead of designing all 16 spacers (four beneath each PCB plus four on the top which close the gap).
Therefore, the thermal resistance had to be simulated according to the real contact characteristics between the PCBs and the spacers.
Moreover, an extra resistance inside the rods located at the height of each PCB, was added to represent the low contact conductance between the threaded rod and both the PCBs and the spacers.

\begin{figure}[!htb]
    \centering
    \includegraphics[width=0.7\columnwidth]{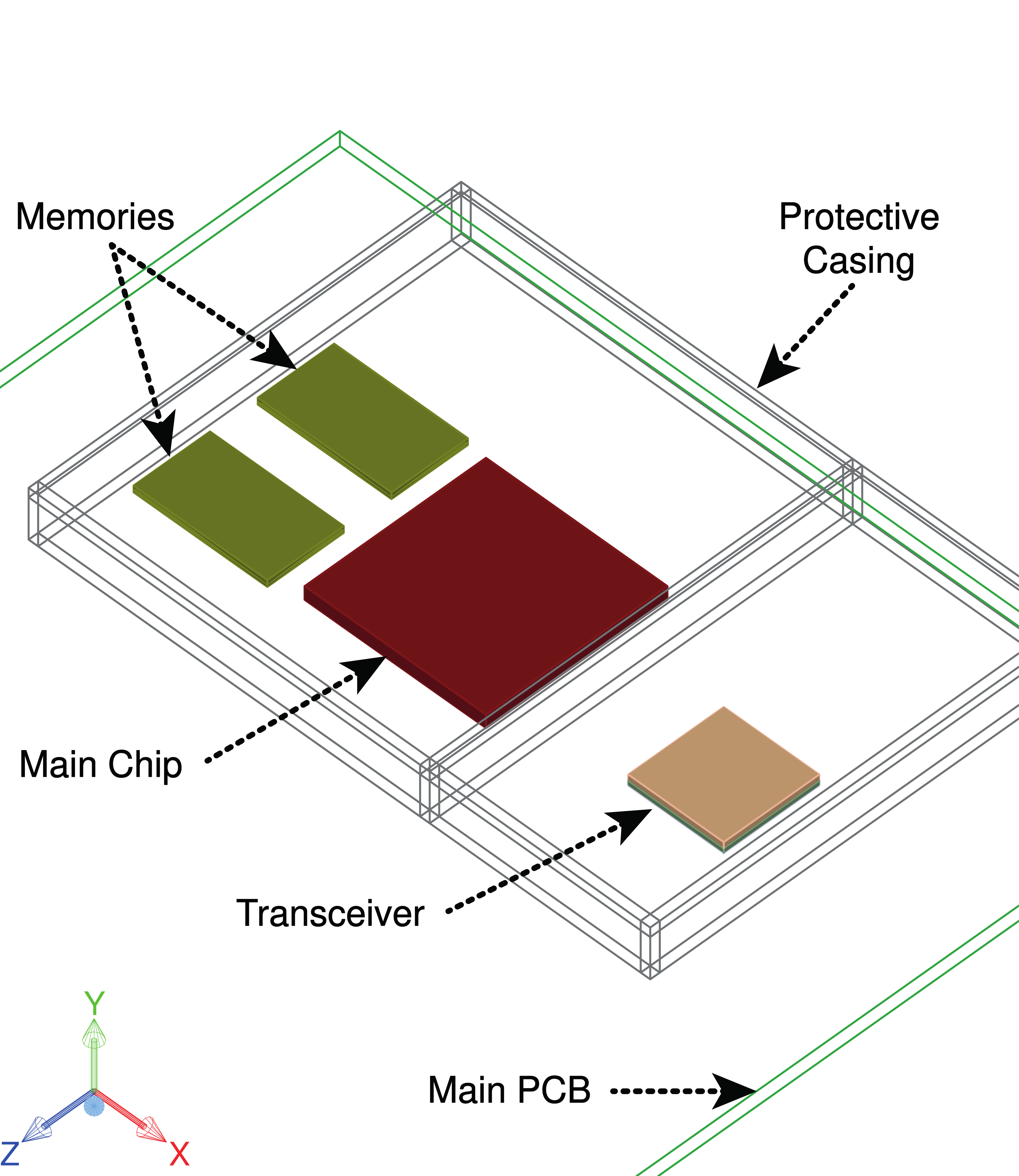}
    \caption{Electronic components mounted on the main PCB and enclosed by an aluminum case.}
    \label{fig:ElectronicComponents}
\end{figure}

The next step is to add the individual models for the following electronic devices: a main chip, two memories, a transceiver, and an amplifier.
While the last one is placed in the RF PCB, the first three components are mounted on the main PCB and enclosed by a protective aluminium case from external radio frequency (RF) interferences, as depicted in Figure~\ref{fig:ElectronicComponents}.
The casing external dimensions are \SI{6.5 x 4.5}{cm}, with walls of \SI{1}{mm} thickness, and it has a height of \SI{0.5}{cm}.
The memories and the main chip are in a separate compartment of the transceiver, divided by a \SI{1}{mm} thick wall.

The structural characteristics of these electronic components, and their positions, are listed in Table~\ref{tab:MainPackages_PosDim}.
While for the main chip, a Xilinx system on a chip of the family Zynq-7000 (model number cl484 xc7z020), there were models available in the Icepak libraries~\cite{Zynq7000_Icepak}, for the others it was necessary to adapt standard packages according to the data sheets of the components.
The memories are commercialised by Micron, model number MT41K256M16TW-107, of the double data rate type three synchronous dynamic random-access memory (DDR3L SDRAM), with \SI{4}{Gb}~\cite{MemoryDataSheet}.
The radio frequency agile transceiver is from Analog Devices, model AD9364~\cite{TransceiverDataSheet}.

\begin{table}[htbp]
    \footnotesize
    \centering
    \caption{Position and characteristics of the electronic packages mounted on the main PCB, with values in \si{cm}.}
    \label{tab:MainPackages_PosDim}
    \begin{tabular}{ccccccc}
        \toprule
            \multirow{2}[0]{*}{Component}   & \multicolumn{3}{c}{Characteristics}   & \multicolumn{3}{c}{Position} \\
        \cmidrule{2-7}
                                            & Width & Length    & Thickness         & X     & Y     & Z \\
        \midrule
        Main Chip                           & 1.90  & 1.90      & 0.14              & 3.64  & \multirow{4}[0]{*}{5.16} & 2.11 \\[1ex]
        Memory 1 & \multirow{2}[0]{*}{1.40} & \multirow{2}[0]{*}{0.80} & \multirow{2}[0]{*}{0.11} & \multirow{2}[0]{*}{1.95} &       & 2.01 \\
        Memory 2                            &       &           &                   &       &       & 3.31 \\[1ex]
        Transceiver                         & 1.00  & 1.00      & 0.15              & 6.71  &       & 2.69 \\
        \bottomrule
    \end{tabular}
\end{table}

\begin{figure}[!htb]
    \centering
    \includegraphics[width=0.8\columnwidth]{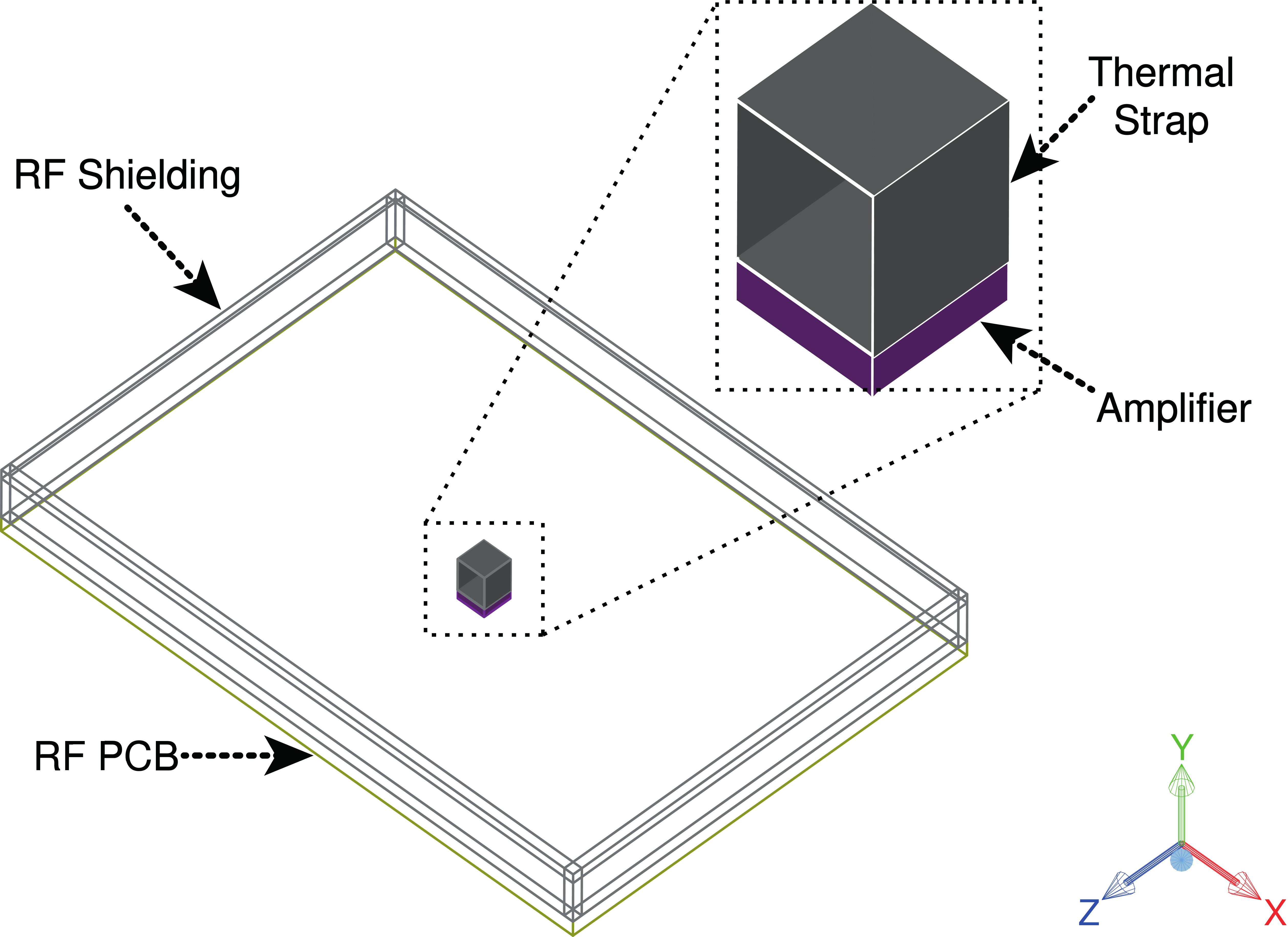}
    \caption{Schematic of the RF PCB and aluminium casing, together with a zoom of the thermal strap and the amplifier.}
    \label{fig:Amplifier}
\end{figure}

As mentioned, the amplifier is mounted alone in the RF PCB board, as depicted in Figure~\ref{fig:Amplifier}.
This package, model number RF5110G, is sold by Qorvo and is a \SI{3}{V} General Purpose, GSM Power Amplifier~\cite{AmplifierDataSheet}.
Since there is no template available of this small amplifier (\SI{0.3 x 0.3 x 0.09}{cm}) for the Icepak software, it was an adaption of a standard one.
To be protected from RF interferences, there is another aluminium case that covers the entire RF PCB board, measuring \SI[allow-number-unit-breaks=true]{6.51 x 4.51 x 0.54}{cm} with \SI{1}{mm} thick walls.
To help to dissipate heat from the amplifier, a thermal strap (TS) has been added between the amplifier and the case, converting the latter into a thermal sink.
This strap is made of pyrolytic graphite, with a thickness of \SI{25}{\mu m}~\cite{PanasonicPGS2017}, and measuring \SI[allow-number-unit-breaks=true]{1.6 x 0.3}{cm} in total.
To maximize heat conduction, the strap is folded in a loop, as is shown in the zoom of Figure~\ref{fig:Amplifier}.

%% file: 3_Analysis.tex
%
%
%


\section{Assumptions and Simulations}
\label{sec:Analysis}

\subsection{Global Assumptions}
\label{subsec:Assumptions}

Several simulations have been run to determine the best configuration, to test different operational modes, and to validate the assumptions for the thermal model.
All of them have been executed both for steady state conditions (giving operational limits values), and for transient conditions (giving temperature evolution values).
For simplicity, only the most relevant simulations and the transient results are described here.

As this is a preliminary analysis and many aspects of the mission are not defined, some major assumptions were made.
The first was the initial temperature, assumed to be at \SI{10}{\degreeCelsius}, which is within what is expected after being launched.

The second major assumption is related to the orbital conditions of the CubeSat, that in this case was assumed to be a Sun-synchronous orbit with an altitude of \SI{400}{km}.
Resorting to the ESA TOPIC software, the worst conditions (in terms of incident fluxes) were searched for within a three years period, and used as input for the walls exterior conditions, since these provide us an educated guess from where design assumptions can be taken.
The orbit also dictated the length of the transient simulations, which had a maximum duration of \SI{5550}{seconds} (about \SI{92.5}{minutes}).

Secondary sensitivity runs were performed to assess the impact of the rods model on the results.
For these runs, some of the parameters that define the rods were modified, in an attempt to better approximate the rods conduction to the real conditions.
The amplifier behaviour was also assessed, in order to better understand its relationship with the operational cycles.

\subsection{Operational Conditions}
\label{subsec:OpConditions}

Most electronic components have a maximum recommended operational temperature of \SI{80}{\degreeCelsius}.
Above this value the efficiency of the component starts to decrease until it reaches the maximum survival temperature (about \SI{120}{\degreeCelsius}).
Therefore, the objective is to make sure that all components are bellow that maximum operational temperature, and that they never reach the survival one.

The power consumption of all components is shown in Table~\ref{tab:PowerDissipated}, for two main distinct cases (when the amplifier is working and when it is not).
As can be seen there, the only two components that do not have a steady power consumption are the amplifier and its PCB (the RF PCB).

\begin{table}[htbp]
    \footnotesize
    \centering
    \caption{Power consumption data for each component, reflecting the operational modes of the system.}
    \label{tab:PowerDissipated}
    \begin{tabular}{ccc}
        \toprule
        \multirow{2}[0]{*}{Object}  & \multicolumn{2}{c}{Power [\si{W}]}  \\
        \cmidrule{2-3}
        & Amp. Off  & Amp. On \\
        \midrule
        Amplifier                   & 0         & 2 \\
        Main Chip                   & \multicolumn{2}{c}{1.8} \\
        Memory 1                    & \multicolumn{2}{c}{0.25} \\
        Memory 2                    & \multicolumn{2}{c}{0.25} \\
        Transceiver                 & \multicolumn{2}{c}{1} \\
        Main PCB                    & \multicolumn{2}{c}{0.5} \\
        Top PCB                     & \multicolumn{2}{c}{0}  \\
        Bottom PCB                  & \multicolumn{2}{c}{1}  \\
        RF PCB                      & 0         & 0.25 \\
        \midrule
        \textbf{Total Power}        & \textbf{4.8}   & \textbf{7.05} \\
        \bottomrule
    \end{tabular}
\end{table}

\subsection{Simulation Options}
\label{subsec:Cases}

\begin{figure}[!htb]
    \centering
    \includegraphics[width=1\columnwidth]{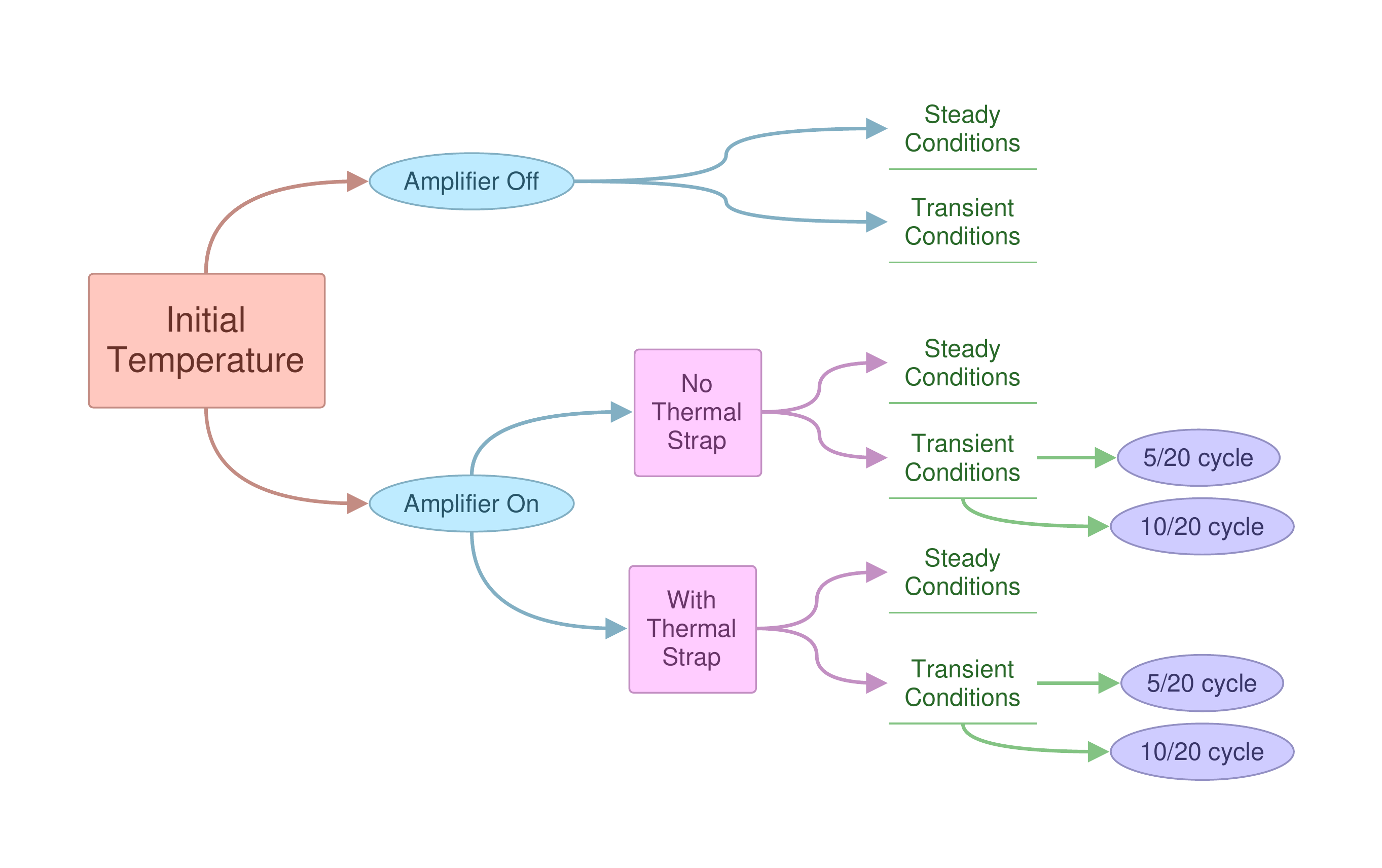}
    \caption{Simplified schematic of the options that have been tested for the CubeSat operations.}
    \label{fig:Sim_Options}
\end{figure}

As is shown in Table~\ref{tab:PowerDissipated}, the amplifier is the electronic component which consumes more power.
Considering its size, it is also the one generating more heat.
However, unlike any other it can be turned off for brief periods.
Therefore, it is the amplifier mode of operation that has the highest impact in the state of the system.
This is translated by the first level of the schematic of the simulated options, represented in Figure~\ref{fig:Sim_Options}.

When the amplifier is turned on, there are still different operational cycles that have to be tested.
Even though the amplifier follows a squared wave pattern, it can either operate in a shorter mode or a longer one.
While for the former the amplifier is active for \SI{5}{seconds}, in the latter mode it works for \SI{10}{seconds}.
Both are followed by an off period of \SI{20}{seconds}.
As this affects transient state conditions, it is represented last in Figure~\ref{fig:Sim_Options}.

An intermediate level was added after having the first simulation results of the temperature distribution.
Because the amplifier was reaching temperatures above the survival limit, a thermal strap was added to the system, in an attempt to mitigate this.
All the above conditions were then tested with and without the thermal strap.
Although several configurations were tested for the strap, the results only reflect the final option (which was the one capable of dissipating more heat), represented in Figure~\ref{fig:Amplifier}.


%% file: 4_Results.tex
%
%
%

\section{Results and Discussion}
\label{sec:Results}

\subsection{Amplifier Off}
\label{sec:AmpOff}

The first step taken was to determine the temperature distribution when the amplifier is turned off, to set a baseline for the system, as this gives figures of merit for temperatures of the main electronic components.

\begin{figure}[!htb]
    \centering
    \includegraphics[width=0.9\columnwidth]{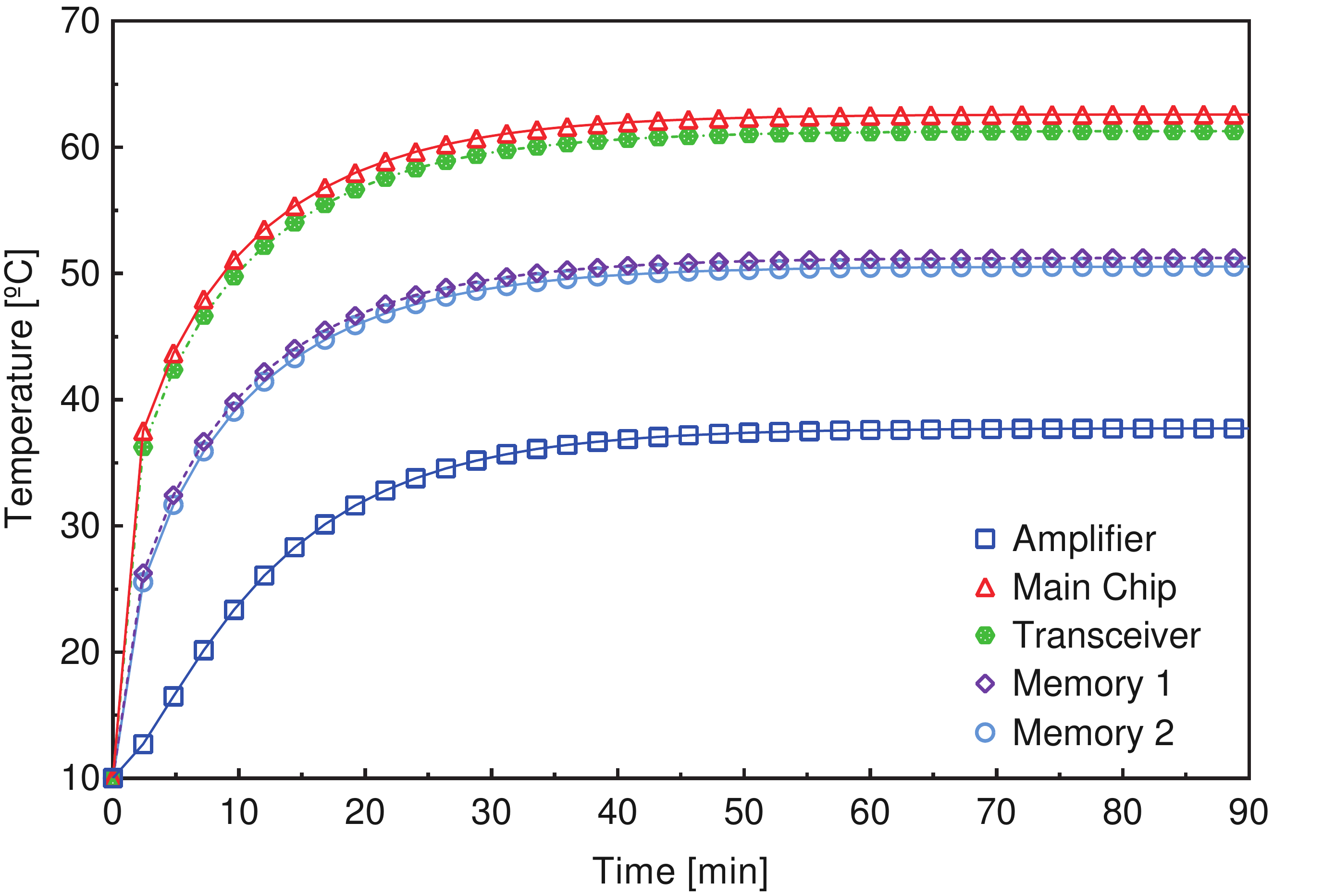}
    \caption{Evolution of the temperature for all electronic components for a full orbital period, when the amplifier is turned off.}
    \label{fig:Graph1_T_AOFF_v1}
\end{figure}

The results obtained are shown in Figure~\ref{fig:Graph1_T_AOFF_v1}, where it is clear that the hottest component is the main chip, reaching a maximum temperature around \SI{63}{\degreeCelsius}.
Not far behind we have the transceiver, with a maximum difference of \SI{4}{degrees}.
Conversely, the memories are about \SI{12}{degrees} colder that the main chip, with a difference between them bellow \SI{1}{degrees}.
Because the amplifier is turned off, the temperature difference to the main chip is even more than the memories (about \SI{28}{degrees}).
After around \SI{33}{minutes} the temperatures of the components stabilise reaching a plateau, with differences bellow \SI{0.01}{degrees}.

From this point forward the main chip is considered a representative of the others, since not only is its temperature evolution very similar to the transceiver, but also it is significantly hotter that the memories.

\subsection{Amplifier On}
\label{sec:AmpOn}

\begin{figure}[!htb]
    \centering
    \includegraphics[width=0.9\columnwidth]{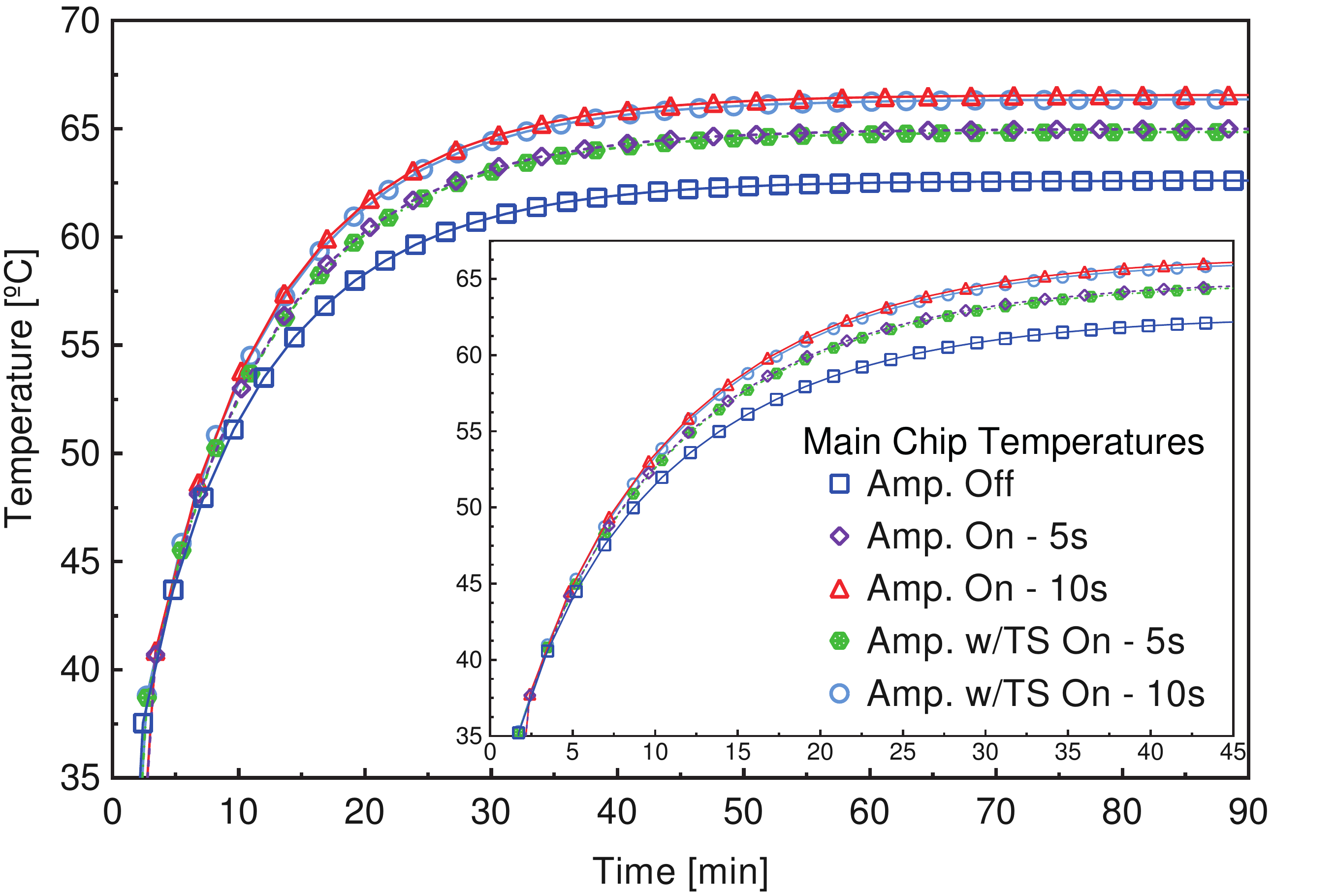}
    \caption{Evolution of the temperature for the main chip in all operational modes and configurations (including a zoom of the first \SI{45}{minutes}).}
    \label{fig:Graph2_1_T_AOn_MainChip_v1}
\end{figure}

From the moment the amplifier is turned on, all components suffer a temperature rise.
Figure~\ref{fig:Graph2_1_T_AOn_MainChip_v1} shows the differences for the main chip, depending on the operational mode of the amplifier and the configuration type (\textit{i.e.} with or without TS).
In the worst case, when the amplifier is working in a \SI{10/20}{seconds} cycle and there is no thermal strap, the temperature rises by up to \SI{4}{degrees}.
The same result is verified in the other components, for the different modes of operation and used configurations.
Nevertheless, this does not hold for the amplifier itself.

When we turn on the amplifier, which is done only after an initial \SI{20}{seconds} hiatus, the temperatures start to rise immediately, reaching values higher than the \SI{80}{\degreeCelsius} limit.
The maximum difference, in comparison when it was turned off, was computed to be \SI{94}{degrees}.

\begin{figure}[!htb]
    \centering
    \includegraphics[width=0.9\columnwidth]{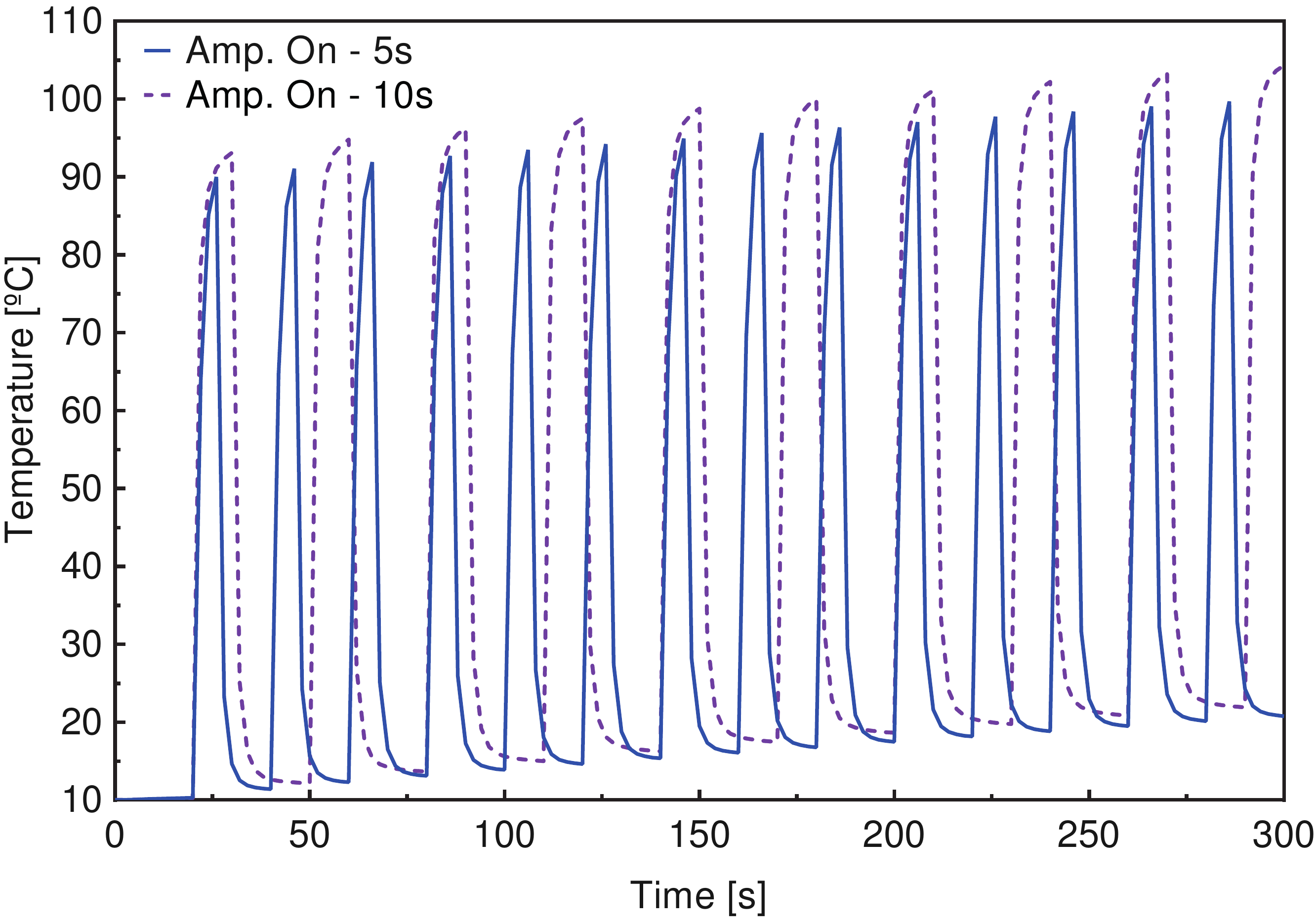}
    \caption{Comparison of the evolution of the temperature for the amplifier operating either with a \SI{5/20}{seconds} cycle or a \SI{10/20}{seconds} one without any thermal strap (only the first \SI{300}{seconds} are shown for better readability).}
    \label{fig:Graph3_T_AOn_Amp5_10_v1}
\end{figure}

We started by comparing the results for different modes of operation of the amplifier and the same configuration (for the first \SI{5}{minutes} of operations).
Analysing the case there is no thermal strap, plotted in Figure~\ref{fig:Graph3_T_AOn_Amp5_10_v1}, it is clear that the \SI{10/20}{seconds} cycle reaches higher temperatures faster (with a maximum temperature difference of about \SI{7}{degrees}).
Nevertheless, both operation cycles are much similar in their warming up and cooling down phases.

\begin{figure}[!htb]
    \centering
    \includegraphics[width=0.9\columnwidth]{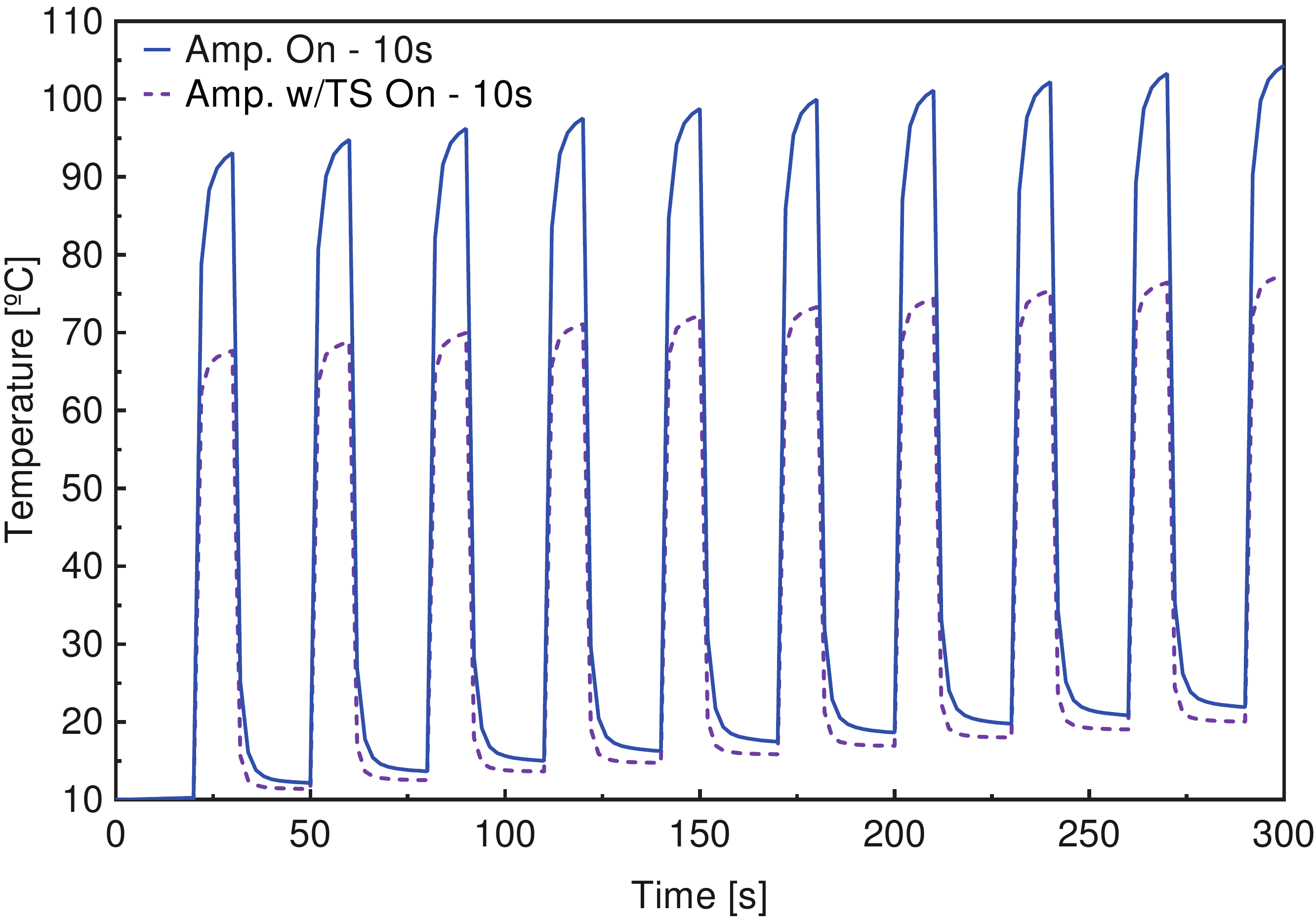}
    \caption{Comparison of the evolution of the temperature for the amplifier with or without the thermal strap and \SI{10/20}{seconds} operational cycle (only the first \SI{300}{seconds} are shown for better readability).}
    \label{fig:Graph4_T_AOn_Amp10_10TS_v1}
\end{figure}

When the thermal strap is added to the model, there is a decrease in the maximum temperature on the amplifier of more than \SI{25}{degrees}, in either mode of operation.
This is clearly reflected in Figure~\ref{fig:Graph4_T_AOn_Amp10_10TS_v1}, for the \SI{10/20}{seconds} cycle, and support the use of the thermal strap.
In fact, both warming up and cooling down phases are almost the same, with the only difference being the temperature reach, which is lower when there is a thermal strap.
This is supports the theory that what is making the temperature drop is the thermal strap presence, and not the operational cycle of the amplifier.

\begin{figure}[!htb]
    \centering
    \includegraphics[width=0.9\columnwidth]{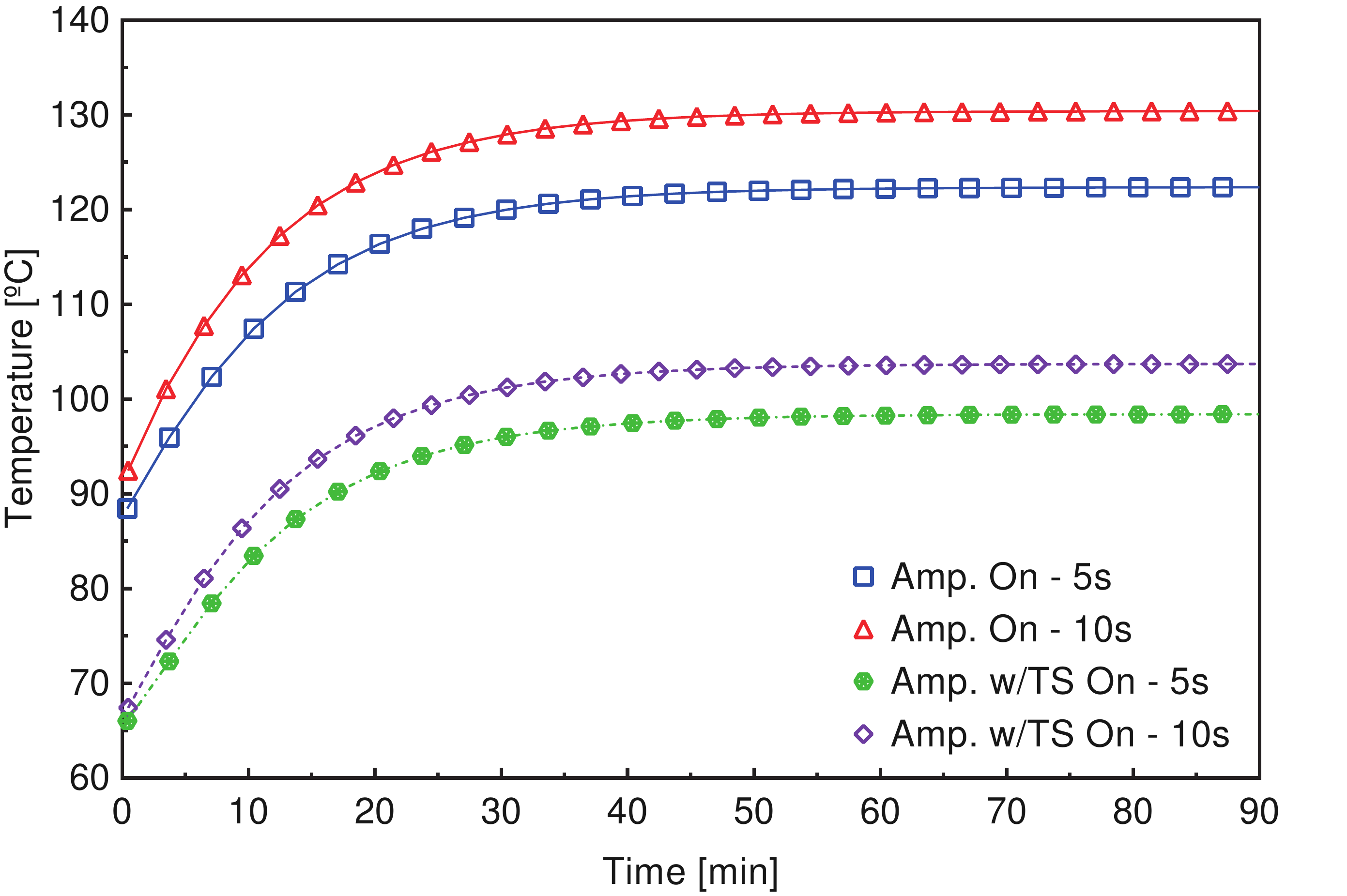}
    \caption{Evolution of the maximum temperature reached by the amplifier for the different operational cycles and with and without the thermal strap.}
    \label{fig:Graph5_T_AOn_AmpComp_v1}
\end{figure}

Extracting from the data just the maximum temperatures reached for each cycle and the different models (with and without the thermal strap), since the phases are similar in all cases, we get the plot of Figure~\ref{fig:Graph5_T_AOn_AmpComp_v1}.
The differences discussed so far for the amplifier are clearly visible with this figure.
For example, for the case the amplifier is working in a \SI{10/20}{seconds} operational cycle, the maximum temperature falls from about \SI{131}{\degreeCelsius}, when there is no thermal strap, to around \SI{103}{\degreeCelsius}.
For the other mode of operation (\SI{5/20}{seconds}), the difference is about \SI{25}{degrees} from a model without to one with a thermal strap (\SI{124}{\degreeCelsius} to \SI{99}{\degreeCelsius}).
Still, the temperatures start to reach higher values than the operational limit of \SI{80}{\degreeCelsius} after \SI{6}{minutes} of operation for the \SI{10/20}{seconds} cycle (and after \SI{7}{minutes} for the \SI{5/20}{seconds} cycle).

\subsection{Temperature Distribution -- Amplifier On}
\label{sec:TempDist_AmpOn}

\begin{figure}[!htb]
    \centering
    \begin{subfigure}[b]{0.7\columnwidth}
        \centering
        \includegraphics[width=1\textwidth]{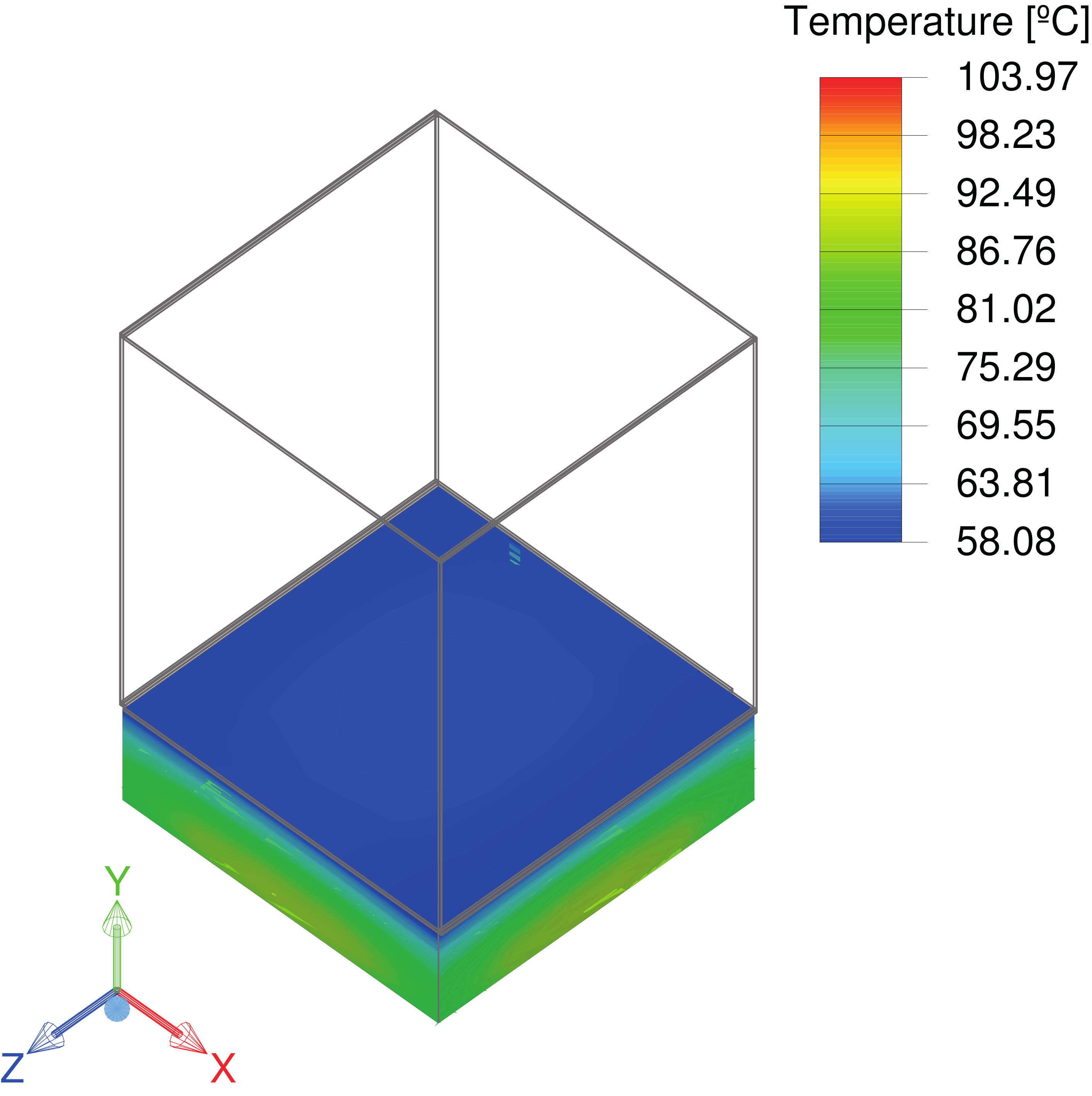}
        \caption{Seen from the top}
        \label{fig:Amplifier_Temp_Front}
    \end{subfigure}%
    \\
    \vskip0.5cm
    \begin{subfigure}[b]{0.7\columnwidth}
        \centering
        \includegraphics[width=1\textwidth]{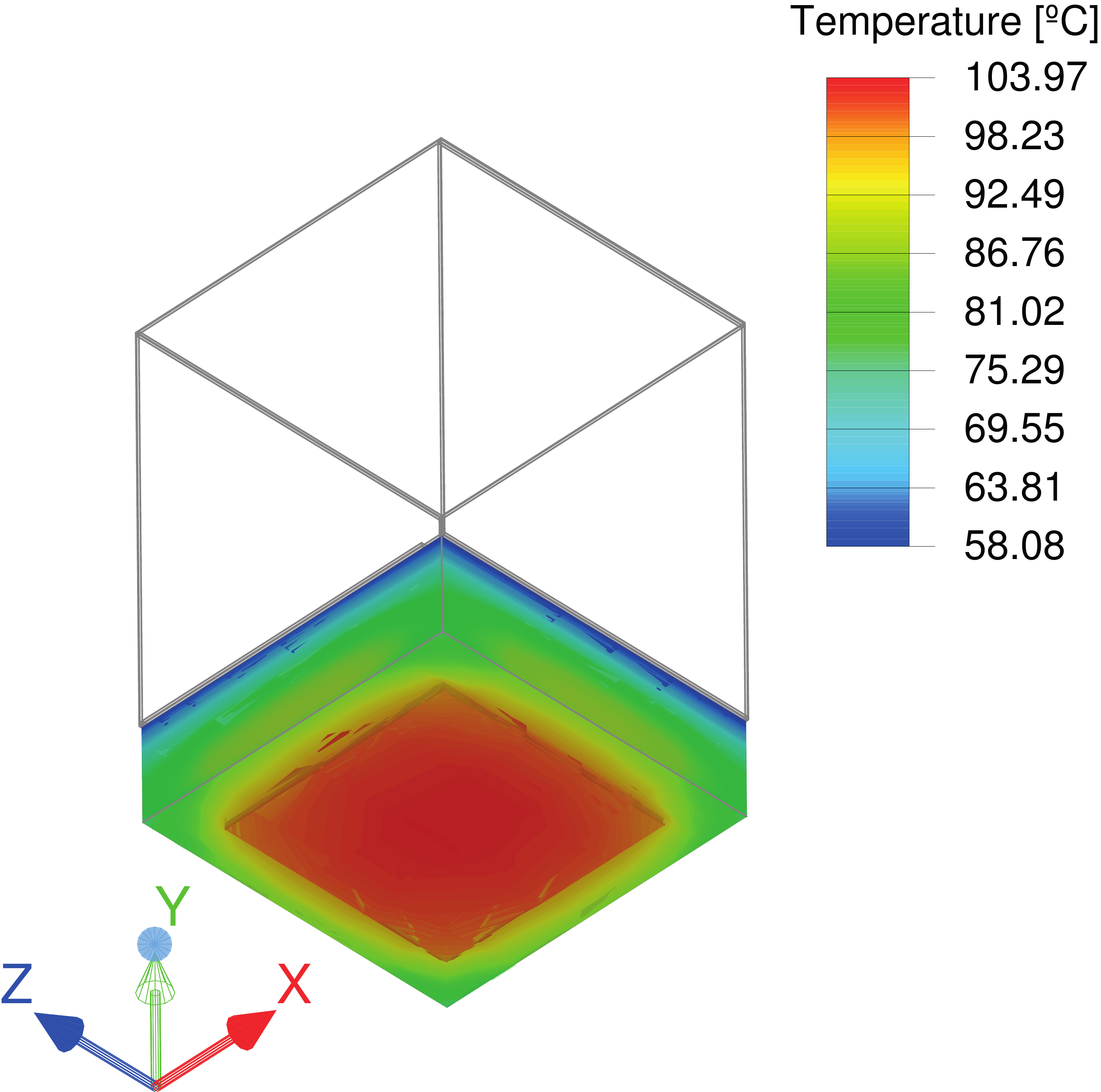}
        \caption{Seen from the bottom}
        \label{fig:Amplifier_Temp_Back}
    \end{subfigure}
    \caption{Temperature distribution on the amplifier after \SI{90}{minutes}.}
    \label{fig:Amplifier_Temp}
\end{figure}

\begin{figure}[!htb]
    \centering
    \includegraphics[width=0.7\columnwidth]{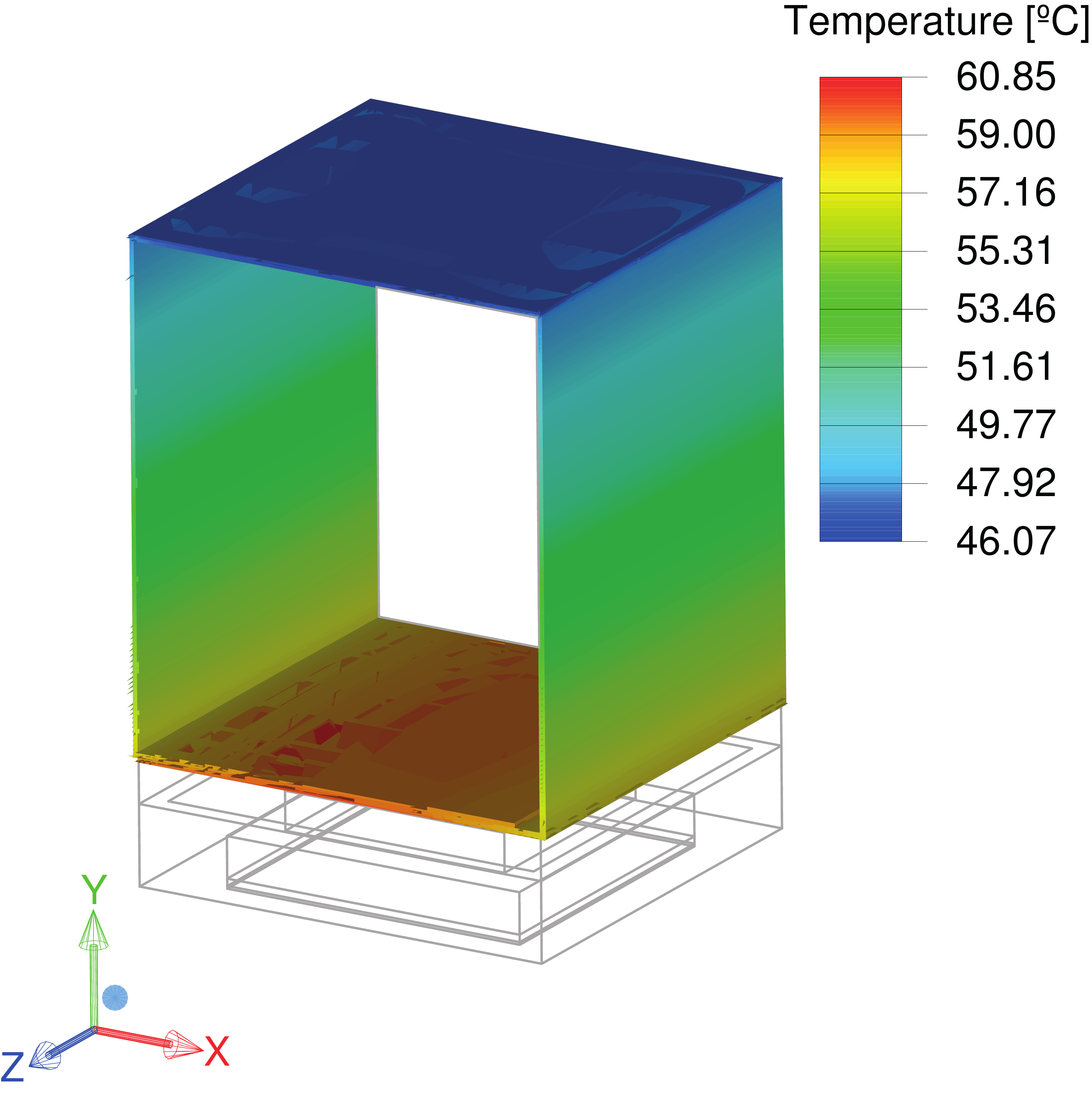}
    \caption{Ttemperature distribution within the thermal strap after \SI{90}{minutes}.}
    \label{fig:ThermalStrap_Temp}
\end{figure}

Although we have to look at the hottest point of each component to understand what are the critical points, as was done above, it is also interesting to see the actual temperature distribution.
For the amplifier, for example, working in a \SI{10/20}{seconds} cycle with the thermal strap present, the temperature distribution is represented in Figure~\ref{fig:Amplifier_Temp}.
This way we can see that the hottest point of the amplifier, at a temperature of about \SI{104}{\degreeCelsius} (shown in Figure~\ref{fig:Amplifier_Temp_Back}), is on the bottom part.
Conversely, the thermal strap only reaches a maximum temperature of approximately \SI{61}{\degreeCelsius} above the amplifier, as seen in Figure~\ref{fig:ThermalStrap_Temp}.

The reason behind these temperatures difference (of about \SI{43}{degrees}) is the distribution inside the amplifier itself, which due to the internal configuration and materials conducts most of the heat to its thermal pad, on the bottom.
Thus, few heat goes to the top, minimising what the thermal strap can actually dissipate.
Nevertheless, it is clear in Figure~\ref{fig:ThermalStrap_Temp} that the strap is conducting similar amounts of heat through its sides.
This was expected due to the loop configuration used, and represents at the end a temperature difference between the bottom and the top of almost \SI{15}{degrees}.

\begin{figure}[!htb]
    \centering
    \begin{subfigure}[b]{1\columnwidth}
        \centering
        \includegraphics[width=1\textwidth]{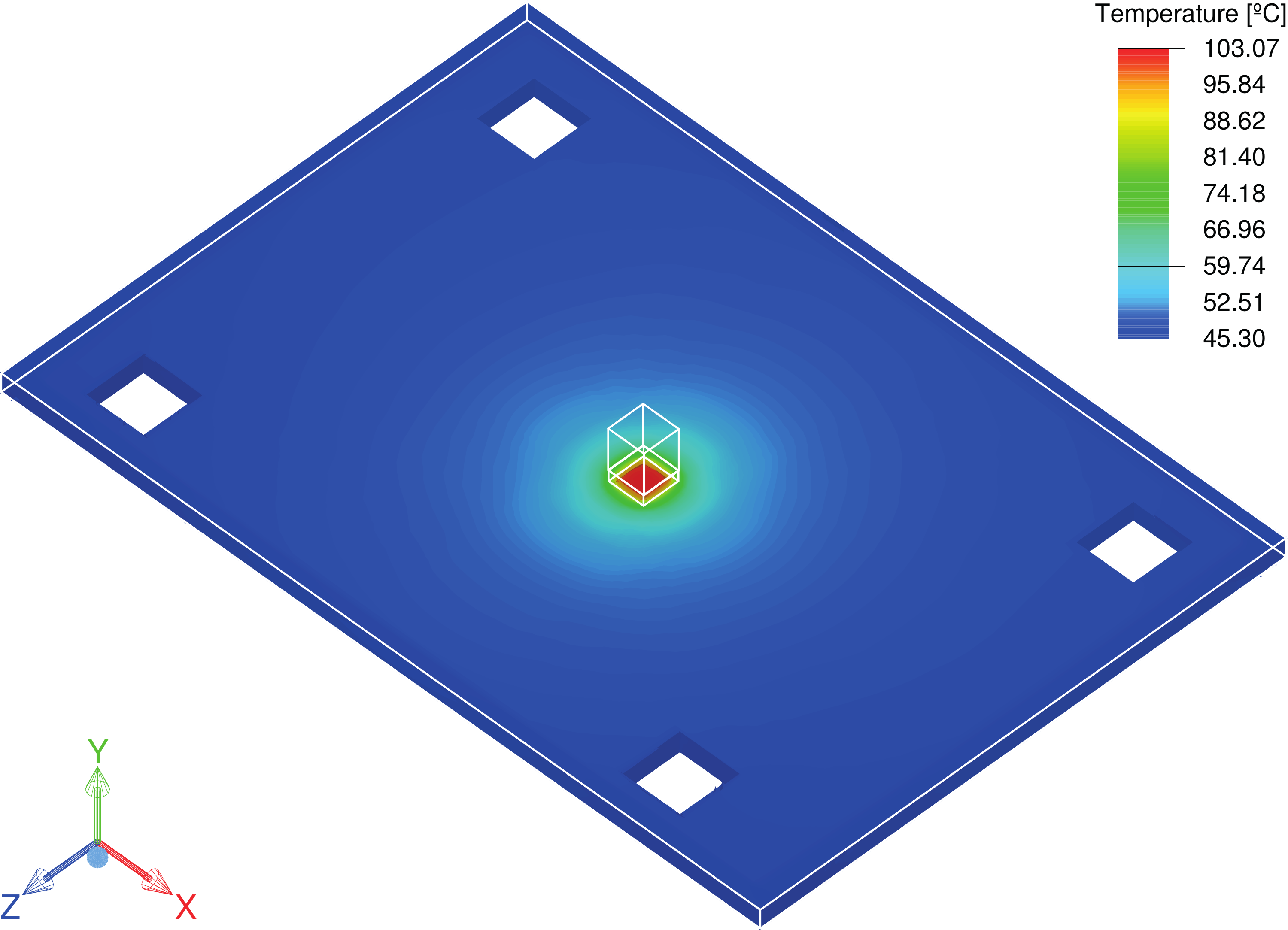}
        \caption{Full view}
        \label{fig:RFPCB_Temp_Full}
    \end{subfigure}%
    \\
    \vskip0.5cm
    \begin{subfigure}[b]{1\columnwidth}
        \centering
        \includegraphics[width=1\textwidth]{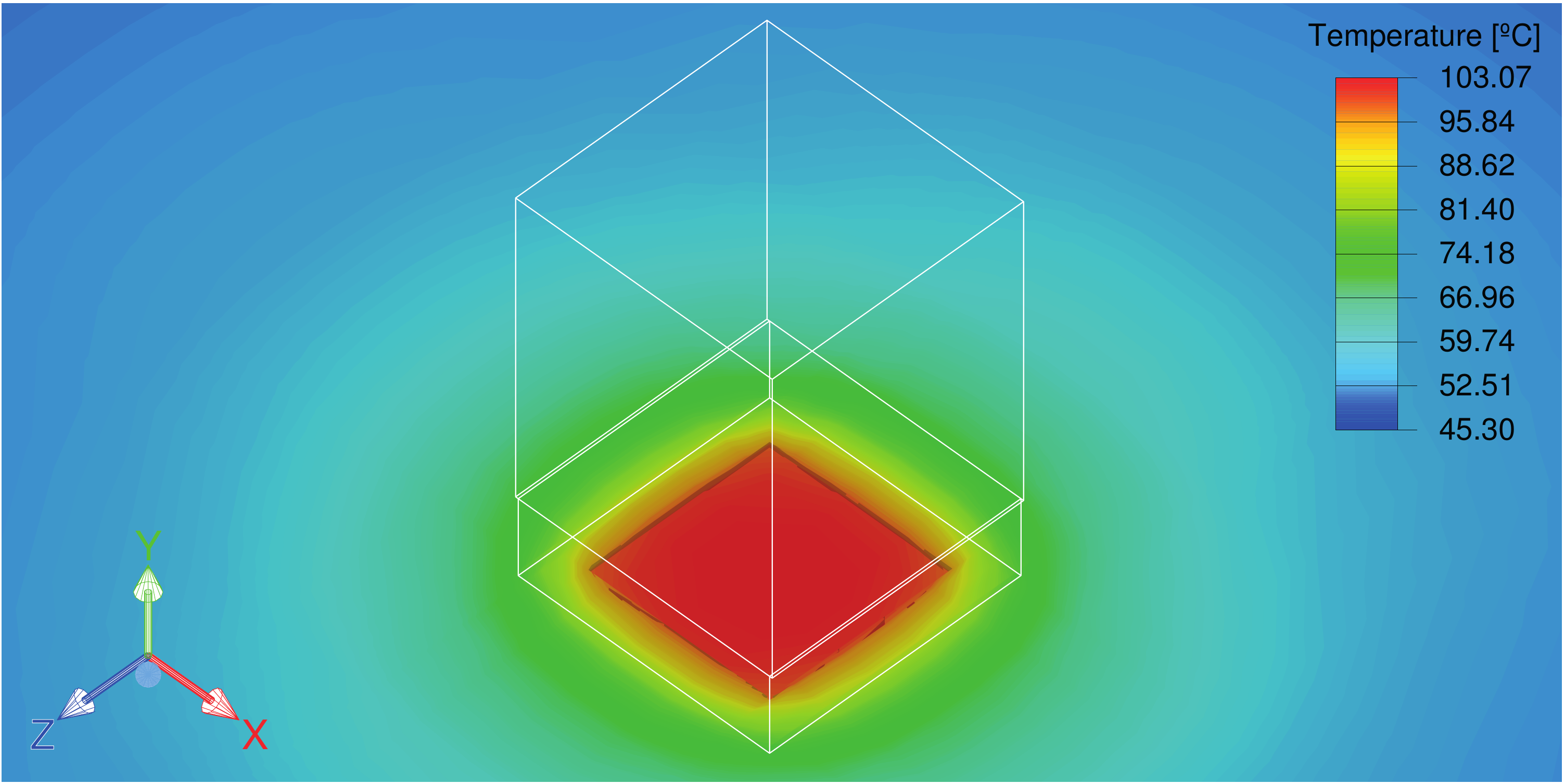}
        \caption{Close up of the amplifier location}
        \label{fig:RFPCB_Temp_CloseUp}
    \end{subfigure}
    \caption{Temperature distribution on the PCB that supports the amplifier (the RF PCB) after \SI{90}{minutes}.}
    \label{fig:RFPCB_Temp}
\end{figure}

Because of the temperature distribution on the amplifier, the RF PCB (which supports it) has its maximum temperature beneath and around the amplifier.
This is depicted in Figure~\ref{fig:RFPCB_Temp}, which also reveals the poor conduction capability of the PCBs, since the rest of the board remains moderately cool.
The few heat that reaches the supporting rods is conducted to the main PCB beneath it.

\begin{figure}[!htb]
    \centering
    \begin{subfigure}[b]{1\columnwidth}
        \centering
        \includegraphics[width=1\textwidth]{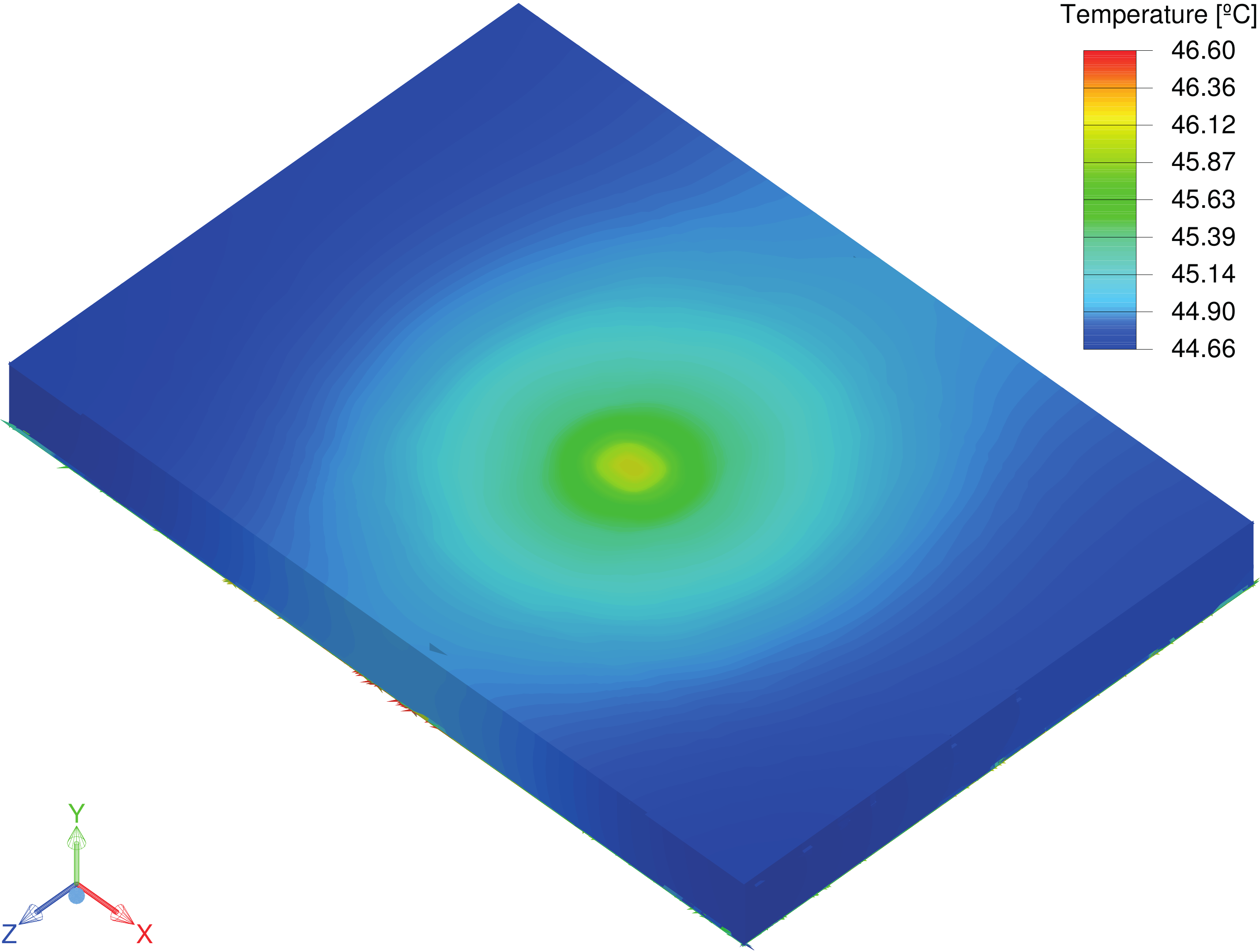}
        \caption{Full shield}
        \label{fig:Shielding_Temp_Full}
    \end{subfigure}%
    \\
    \vskip0.5cm
    \begin{subfigure}[b]{1\columnwidth}
        \centering
        \includegraphics[width=1\textwidth]{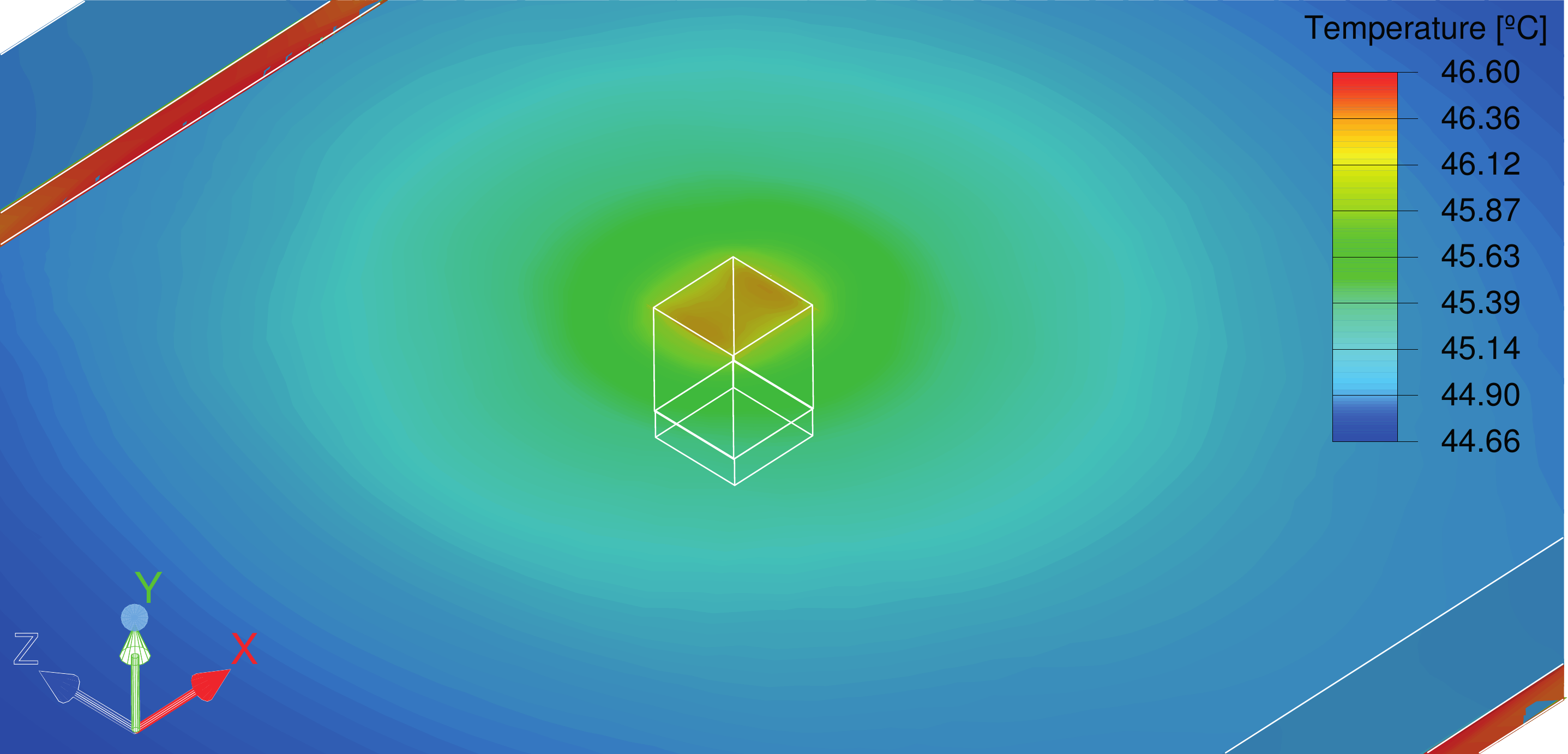}
        \caption{Close up to the amplifier location}
        \label{fig:Shielding_Temp_CloseUp}
    \end{subfigure}
    \caption{Temperature distribution on the shield casing of the amplifier after \SI{90}{minutes}.}
    \label{fig:Shielding_Temp}
\end{figure}

Conversely, the casing, which is linked to the amplifier by the thermal strap, is just moderately warmed (as is show in Figure~\ref{fig:Shielding_Temp}).
The maximum temperatures of the casing are observed close to where the thermal strap is attached to and where the casing connects to the RF PCB.
Nonetheless, it seems to be capable of conducting more heat than what it does now, since the temperature range shown is lower than \SI{2}{degrees}.

%% file: 6_Conclusion.tex
%
%
%

\section{Conclusions and Future Work}
\label{sec:Conclusion}

The results of the computational analyses performed revealed that the temperature of the amplifier are out of its long-term sustainable range.
The impact of this issue in the overall mission would most likely cause the loss of the amplifier or, at least, a degradation in its performance and lifetime.
Therefore, the thermal subsystem should not be underrated when designing any type of spacecraft, from small to large conventional satellites.
However, thermal analysis are not without limitations, with several parameters having to be estimated.
Therefore, not only design safety margins must be accounted for, but also several tests should be performed on single components, system testbeds and the full spacecraft, before it is launched.

When the amplifier is operational, there is a clear impact on the temperatures off all components, despite the geometry of the system (\textit{i.e.} with or without a thermal strap) or the operational mode of the amplifier (\SI{5/20}{seconds} or \SI{10/20}{seconds} cycles).
While the main chip was the component showing higher temperatures before, once the amplifier is turned on its temperature goes well above the ones from the other components, reaching in some cases \SI{131}{\degreeCelsius}.

The implementation of a thermal strap to conduct the heat away from the amplifier is fundamental, with temperatures falling up to \SI{27}{degrees}.
Conversely, the change in operational mode does not produce a significant impact on the temperatures (with maximum differences of just up to \SI{7}{degrees}).
With shorter cycle operations the \SI{80}{\degreeCelsius} limit is reached within \SI{1}{minute}.

\section{Future Work}
\label{sec:Future_Work}

Regardless of operational mode or configuration, the amplifier still exceed the recommended operational temperature limit.
Even though this is considering that the amplifier would work for an entire orbital period, solutions have to be devised to minimise this problem.

Looking at the temperature distribution on the amplifier it is apparent that the hotter areas are at the bottom of the component.
Moreover, the PCB that supports it does not conduct enough heat to cool the amplifier, while the casing on top is still only moderately warm.
Thus, a possible solution would be to encapsulate the entire amplifier, and the surrounding PCB area, with the thermal strap.
This would transport more heat to the casing, and away from the amplifier, while overcoming the heat distribution limitations arising from the amplifier internal arrangement.

Another solution would be to add an extra thermal strap to the bottom of the PCB.
For this a different heat sink would have to be selected, considering that it has to be a good heat conductor.
This invalidates the use of the main PCB beneath it, due to their poor conduction capabilities.
At the same time, it is important not to add too much stiffness to the model, because of the PGS thermal strap fragility.
Furthermore, this new heat path must have proper electrical isolation so that short circuits are avoided.

If the mission proceeds to the next phase, new and more detailed analyses should be build and run.
In particular, these new simulations should include all internal components (with a proper distribution and the materials selected), and account for all internal power dissipations.
Furthermore, the spacecraft orbital data should be assessed, as well as the thermal environment.

Apart from new simulations, it is also relevant to run laboratory tests on some components.
A particular important assumption made, and that should be validated, is the conduction between the PCBs stack and the rods that support it.
